\documentclass[sigconf]{acmart}

\microtypesetup{expansion=false}

\AtBeginDocument{%
  }

\usepackage{tikz}
\usetikzlibrary{arrows.meta, positioning, calc}

\usepackage{latexml}

\usepackage{array}

\copyrightyear{2026}
\acmYear{2026}
\setcopyright{cc}
\setcctype{by}
\acmConference[ICAIF '26]{7th ACM International Conference on AI in Finance}{November 14--17, 2026}{Milan, Italy}
\acmBooktitle{7th ACM International Conference on AI in Finance (ICAIF '26), November 14--17, 2026, Milan, Italy}
\acmDOI{10.1145/3858063.3858894}
\acmISBN{979-8-4007-3087-0/2026/11}
\begin{document}
%% Compound-word-heavy prose + expansion=false overfills lines; give TeX slack.
\emergencystretch=1.5em

\title[Amortized Neural Filtering for Calibrated Posteriors]{Amortized Neural Filtering for Calibrated Posteriors in Statistical Arbitrage}

\author{Minkey Chang}
\orcid{0009-0001-8175-2583}
\affiliation{%
  \institution{Seoul National University}
  \city{Seoul}
  \country{Republic of Korea}}
\email{minkey.chang@gmail.com}

\renewcommand{\shortauthors}{Chang}

\begin{abstract}
Statistical arbitrage trades on an estimate of the state of a
residual spread, and conventional practice discards the uncertainty
of that estimate. We keep the uncertainty and use it in the trading decision. A Neural Filter Estimator (NFE) is trained on simulated paths of a mean-reverting process whose
innovation law is selected by testing candidates against real
residuals; once trained, it
returns the posterior over the latent state of every spread in a
cross-section in one forward pass. Three results follow. On simulated
paths with known truth, every practical estimator, ours included,
falls in one narrow band of accuracy well below an estimator given
full information, so the posterior gains no accuracy over a point
estimate; its contribution is a correct account of its own
uncertainty. On Korean equities, net of costs, the conventional point
estimate forecasts about a thousand basis points per holding period more than it
realizes, and the amortized methods cut that gap by a factor of
roughly two to eight. Sizing positions from the posterior on the same
learned factors as the end-to-end Attention Factor Model matches the
Sharpe ratio of those factors under the base rule within the period's
noise, at no more risk.
\end{abstract}

%% CCS concepts.


\maketitle

\section{Introduction}
\label{sec:introduction}

In the old tale of the blind men and the elephant, each man touches
the animal and describes a different beast. Economics and finance
pose the same difficulty. It is nearly impossible to say where the
market stands \emph{at the moment}, and certainty arrives only after
the fact. A familiar example is the business cycle, whose recessions are
dated by the NBER many months after they begin \cite{nber-bcdc}. The
term \emph{nowcasting} arose from this predicament
\cite{giannone-2008-nowcasting}. Rather than wait
for the whole animal to come into view, we estimate the \emph{present}
state from the part within reach. A forecast means little without
knowing where one stands. The same problem appears in statistical
arbitrage, where knowing what is happening now matters more to the
next action than predicting what happens next.

Conventional statistical arbitrage \cite{gatev-2006-pairs,
avellaneda-2010-statarb} handles this problem with a hand-crafted
pipeline. One regresses returns on factors, accumulates the
residuals, fits a mean-reversion model on a trailing window, and
uses the fitted point estimates to decide when to enter and exit.
The procedure treats the assumed process as if it were true, and
despite the name \emph{statistical} arbitrage it discards the
uncertainty of the estimates, so that a genuine opportunity and an
optimistic misreading of the data become indistinguishable. In
practice, moreover, the procedure is applied to a large universe of
candidate spreads: the candidates are ranked by these estimates and
the top-ranked ones are traded. However, such ranking compounds the
problem, because the candidates with the most favorable estimation
errors are precisely the ones selected; this is the winner's curse
of post-selection inference \cite{efron-2011-tweedie}. More recent
work learns the factors end to end \cite{guijarro-2021-deepstatarb,
epstein-2025-attentionfactors}. This improves the construction of
the residuals but leaves their use unchanged: the trading decision
is still made from a point estimate, with no estimate of the latent
state and no measure of its uncertainty.

Despite these differences, the methods above share one structure.
Each compresses a panel of returns, one day's cross-section of many
assets, into a few factors, and trades what the factors do not
explain. They differ only in how the factors are obtained.
Fama--French \cite{fama-french-1993} fix them in advance as
characteristic portfolios; Avellaneda--Lee
\cite{avellaneda-2010-statarb} extract them from the data by
principal components and trade the residual spread as a
mean-reverting process; Epstein et al.\
\cite{epstein-2025-attentionfactors} learn them end to end from
characteristics with a trading objective. Outside finance, tabular foundation models \cite{qu-2025-tabicl,
kong-2026-tabfm} perform the same compression with a network
pretrained on a synthetic prior rather than fitted to each dataset. This paper takes the next step in that
sequence. It keeps the residual spread and its mean-reversion state
from Avellaneda--Lee, the task-driven factor learning of Epstein et
al., and the synthetic-prior pretraining of the tabular models, and
it adds what none of them has: a dynamic model of the latent state,
with the uncertainty of its filtered estimate. We call the resulting
estimator the Neural Filter Estimator (NFE).

Our claim concerns how the regime is defined and estimated. Most
statistical-arbitrage strategies already condition on some notion of
regime, but in practice that notion is a heuristic, such as a
volatility threshold or a trailing return, or a label assigned ex post, once the outcome is known. We instead define the regime as a latent variable of an
explicit generative model of the spread: whether the spread is
currently converging to its level or diverging from it, and how
fast. We then show that a single amortized filter can estimate this
regime for every spread in a cross-section from past data alone,
with a calibrated probability attached, and that the trading
decision can use that probability in place of a point estimate.

Two choices in this definition need justification: the number of
states and the use of a probability. In the regime-switching
literature \cite{hamilton-1989-regime}, a latent state is called a
regime when a change in it changes the distribution of the observed
process, for example its mean, its variance or its speed of mean
reversion. For a trader, only the changes that alter the trading
decision matter, so we define the regime relative to the decision.
For a contrarian strategy held over a fixed horizon this leaves two
states: one in which enough of the deviation reverts within the
horizon for a position taken against it to pay, and one in which it
does not, so that the same position earns nothing but its costs.
Acting on this regime requires a judgment of how certain its
assignment is, because the two states call for different actions
and a mistaken assignment is costly. A point estimate assigns the spread
to one state outright and gives no measure of that certainty, which
makes the assignment difficult to act on; a posterior assigns a
probability to each state, which the decision can weigh against the
cost of being on the wrong side. Conventional practice lacks this
measure not out of neglect but because its plug-in estimates do not
provide it.

The rest of the paper supports this claim with three measurements.
The first checks the simulator. The filter is trained only on
simulated spreads, so the simulated process must resemble the real
residuals before anything learned from it can be trusted on real
data. We test this directly: a classifier is trained to distinguish
simulated windows from real ones, and a candidate process passes
only if the classifier largely fails. The Gaussian noise of the
classical model is rejected. The only candidate that passes on both cross-sections has both
time-varying volatility and asymmetric shocks; fat tails alone are
not enough.

The second measures how well the regime can be identified at all.
On simulated spreads the true regime is known, so every method can
be compared with the best possible classifier, one that knows the
true model, level and volatility. Every practical method, ours
included, lands in one narrow band far below that ideal. The
shortfall comes not from the estimator but from the window: the
level of the spread must be estimated from the same window as its
reversion, and that alone accounts for most of the loss. Better
inference cannot close this gap; only more information can.

The third asks what the posterior changes in trading. Under fixed
trading rules and after costs, posterior sizing earns about the same as the point estimate; the
posterior adds no alpha. What changes is the
strategy's own forecast of its return. The point estimate predicts
far more than it realizes, a direct consequence of the winner's
curse above; the posterior-based strategies reduce this gap
several-fold. The mean-reverting structure in the data is real, but
it is smaller than the cost of trading on it, and the value of the
posterior lies in saying so.

\section{Related Work}
\label{sec:related}

\emph{Factor models and statistical arbitrage.} The line of work this
paper extends begins with static characteristic factors: Fama--French
\cite{fama-french-1993} explain the cross-section of returns with a
few factor portfolios fixed in advance. Dynamic factor models let the
factors evolve as a latent linear state \cite{geweke-1977-dfm,
stock-watson-2002-dfm}. Statistical arbitrage trades the residuals
that the factor models leave unexplained. Gatev et al.\
\cite{gatev-2006-pairs} trade pairs selected by the distance method,
and Avellaneda and Lee \cite{avellaneda-2010-statarb}, in what has
become the standard treatment, fit a mean-reversion model to
cumulative factor residuals and trade its point estimates. Recent
deep-learning work trades factor portfolios instead of selected pairs
and learns the trading policy from residual histories end to end
\cite{guijarro-2021-deepstatarb}, and
\citet{epstein-2025-attentionfactors} learn conditional factors by
attention over characteristics, jointly with an after-cost trading
objective. In all of these methods the trading decision rests on a
point estimate, and the uncertainty of the latent state plays no
part in it.

\emph{Tabular foundation models.} Outside finance, a parallel line
of work arrived at the same computation. Prior-data fitted networks \cite{muller-2022-pfn} train a
transformer on draws from a synthetic prior, so that one forward pass
approximates a posterior predictive. \citet{hollmann-2023-tabpfn}
make the approach practical for small tables with TabPFN
\cite{hollmann-2025-tabpfn}, \citet{qu-2025-tabicl} scale it with a
column-then-row encoding, and \citet{kong-2026-tabfm} combine these
in TabFM, which alternates attention across columns and rows,
compresses each row to one vector, and predicts from those vectors
in context. A
factor model also compresses each cross-section into a few numbers;
in this sense the conditional factor of
\citet{epstein-2025-attentionfactors} is the finance special case of
the learned row embedding. These models are not built for financial
panels: their in-context stage attends over rows as an exchangeable set, and
a time-ordered panel is not one. This paper closes part of that gap
by placing a latent-state dynamic model under the row embedding and
inferring the state from the past alone.

\emph{Filtering and amortized inference.} In state-space models,
filtering refers to the estimation of a latent state from past
observations alone \cite{kalman-1960-filter}. Hamilton
\cite{hamilton-1989-regime} introduced its regime-switching form to
economics, and modern deep variants learn the state-space model
itself \cite{krishnan-2015-dkf} or amortize the filter with a
variational inference network \cite{marino-2018-avf}. When the
latent state can be simulated together with the observations, a different approach becomes possible. Neural posterior estimation
trains a conditional density estimator on simulator draws and then
evaluates the posterior in one forward pass
\cite{papamakarios-2016-npe}; it is the core of simulation-based
inference \cite{cranmer-2020-sbi}, and its calibration can be
checked by simulation \cite{talts-2018-sbc}. The NFE applies neural
posterior estimation to the spreads of a time-ordered panel, restricts
the prior to a dynamic process whose innovation law is tested against
real data, and amortizes the filtering posterior over the state.

\section{Method}
\label{sec:method}

\begin{figure*}[t]
  \centering
  %% LaTeXML renders TikZ poorly, so the HTML version shows the same
  %% drawing as a vector image (pipeline_html.pdf, cut from the TeX build).
  \iflatexml
  \includegraphics[width=\textwidth]{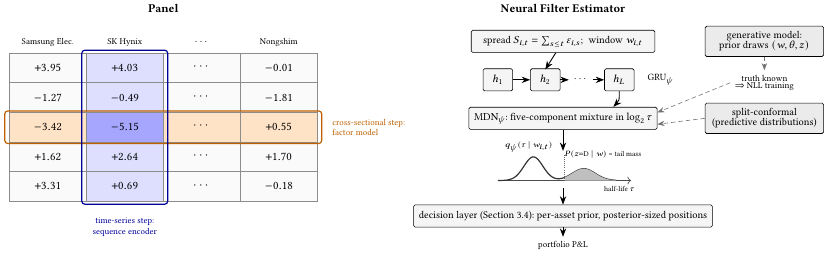}%
  \else
  \resizebox{0.78\textwidth}{!}{%
  \begin{tikzpicture}[
      font=\footnotesize,
      cell/.style={draw=black!45, fill=black!3, minimum width=13pt,
                   minimum height=10pt, inner sep=0pt},
      blk/.style={draw, fill=black!4, rounded corners=1pt, align=center,
                  inner sep=3.5pt, minimum height=15pt},
      lbl/.style={font=\scriptsize, align=center},
      arr/.style={->, >=Stealth, semithick},
      darr/.style={->, >=Stealth, densely dashed, black!55}]

    % ============ LEFT PANEL: the panel, worked example ============
    \begin{scope}[shift={(-268pt, -26pt)},
        xcell/.style={draw=black!45, fill=black!2, minimum width=52pt,
                      minimum height=20pt, inner sep=0pt,
                      font=\small}]
      \node[font=\small\bfseries] at (78pt, 40pt) {Panel};
      % column headers: three KRX stocks
      \node[lbl] at (0pt, 17pt) {Samsung Elec.};
      \node[lbl] at (52pt, 17pt) {SK Hynix};
      \node[lbl] at (104pt, 17pt) {$\cdots$};
      \node[lbl] at (156pt, 17pt) {Nongshim};
      % cells: PCA-3 residual returns in percent, five consecutive days
      \node[xcell] at (0pt, 0pt) {$+3.95$};
      \node[xcell, fill=blue!13] at (52pt, 0pt) {$+4.03$};
      \node[xcell] at (104pt, 0pt) {$\cdots$};
      \node[xcell] at (156pt, 0pt) {$-0.01$};
      \node[xcell] at (0pt, -20pt) {$-1.27$};
      \node[xcell, fill=blue!13] at (52pt, -20pt) {$-0.49$};
      \node[xcell] at (104pt, -20pt) {$\cdots$};
      \node[xcell] at (156pt, -20pt) {$-1.81$};
      \node[xcell, fill=orange!22] at (0pt, -40pt) {$-3.42$};
      \node[xcell, fill=blue!35] at (52pt, -40pt) {$-5.15$};
      \node[xcell, fill=orange!22] at (104pt, -40pt) {$\cdots$};
      \node[xcell, fill=orange!22] at (156pt, -40pt) {$+0.55$};
      \node[xcell] at (0pt, -60pt) {$+1.62$};
      \node[xcell, fill=blue!13] at (52pt, -60pt) {$+2.64$};
      \node[xcell] at (104pt, -60pt) {$\cdots$};
      \node[xcell] at (156pt, -60pt) {$+1.70$};
      \node[xcell] at (0pt, -80pt) {$+3.31$};
      \node[xcell, fill=blue!13] at (52pt, -80pt) {$+0.69$};
      \node[xcell] at (104pt, -80pt) {$\cdots$};
      \node[xcell] at (156pt, -80pt) {$-0.18$};
      % band outlines
      \draw[orange!75!black, rounded corners=2pt, thick]
        (-29pt, -29pt) rectangle (185pt, -51pt);
      \draw[blue!60!black, rounded corners=2pt, thick]
        (23pt, 12pt) rectangle (81pt, -92pt);
      % step labels
      \node[lbl, text=orange!75!black, align=left, anchor=west]
        at (189pt, -40pt) {cross-sectional step:\\factor model};
      \node[lbl, text=blue!60!black, align=center, anchor=north]
        at (52pt, -98pt) {time-series step:\\sequence encoder};
    \end{scope}

    % ================= RIGHT PANEL: the filter =================
    \begin{scope}[shift={(20pt, 12pt)}]
      \node[font=\small\bfseries] at (60pt, 2pt)
        {Neural Filter Estimator};
      \node[blk, minimum width=124pt] (win) at (60pt, -20pt)
        {spread $S_{i,t} = \sum_{s \le t}\varepsilon_{i,s}$;\;
         window $w_{i,t}$};
      \node[blk, minimum width=20pt] (g1) at (16pt, -46pt) {$h_1$};
      \node[blk, minimum width=20pt] (g2) at (48pt, -46pt) {$h_2$};
      \node[lbl] (gd) at (72pt, -46pt) {$\cdots$};
      \node[blk, minimum width=20pt] (gL) at (98pt, -46pt) {$h_L$};
      \draw[arr] (g1) -- (g2); \draw[arr] (g2) -- (gd);
      \draw[arr] (gd) -- (gL);
      \node[lbl, anchor=west] at (114pt, -46pt)
        {$\operatorname{GRU}_\psi$};
      \node[blk, minimum width=124pt] (mdn) at (60pt, -72pt)
        {$\operatorname{MDN}_\psi$: five-component mixture in $\log_2 \tau$};
      \begin{scope}[shift={(16pt, -114pt)}]
        \draw[black!70, ->] (0pt, 0pt) -- (90pt, 0pt);
        \node[font=\tiny, anchor=north] at (82pt, -1pt)
          {half-life $\tau$};
        \draw[thick, black!80, smooth, samples=60, domain=0:80,
              x=1pt, y=1pt]
          plot ({\x}, {16*exp(-((\x-24)/9)^2)
                        + 8*exp(-((\x-58)/11)^2)});
        \fill[black!25, smooth, samples=30, domain=45:80,
              x=1pt, y=1pt]
          plot ({\x}, {16*exp(-((\x-24)/9)^2)
                        + 8*exp(-((\x-58)/11)^2)})
          -- (80pt, 0pt) -- (45pt, 0pt) -- cycle;
        \draw[black!60, densely dotted] (45pt, 0pt) -- (45pt, 17pt);
        \node[font=\tiny, anchor=south west] at (42pt, 12pt)
          {$P(z{=}\textup{D}\mid w)$ = tail mass};
        \node[font=\tiny, anchor=south] at (21pt, 16pt)
          {$q_\psi(\tau \mid w_{i,t})$};
      \end{scope}
      \node[blk, minimum width=124pt] (dec) at (60pt, -138pt)
        {decision layer (Section~\ref{sec:method:decision}):
         per-asset prior, posterior-sized positions};
      \node[lbl] (book) at (60pt, -158pt) {portfolio P\&L};
      \draw[arr] (win) -- (48pt, -39pt);
      \draw[arr] (gL.south) .. controls (98pt, -60pt) ..
        ([xshift=34pt]mdn.north);
      \draw[arr] (mdn) -- (60pt, -94pt);
      \draw[arr] (60pt, -119pt) -- (dec);
      \draw[arr] (dec) -- (book);
      \node[blk, fill=black!8, minimum width=80pt] (sim)
        at (196pt, -20pt) {generative model:\\[-1pt]
        prior draws $(w, \theta, z)$};
      \node[lbl] (nll) at (196pt, -48pt)
        {truth known\\[-2pt] $\Rightarrow$ NLL training};
      \node[blk, fill=black!8, minimum width=80pt] (conf)
        at (196pt, -72pt) {split-conformal\\[-1pt] (predictive distributions)};
      \draw[darr] (sim) -- (nll);
      \draw[darr] (nll.west) -- ([yshift=4pt]mdn.east);
      \draw[darr] (conf.west) -- ([yshift=-4pt]mdn.east);
    \end{scope}
  \end{tikzpicture}}%
  \fi
  \Description{Left: a panel table whose columns are KRX stocks and
    whose rows are consecutive days, with one day's row and one asset's column highlighted. Right: the filter flow from spread window through
    a recurrent encoder and mixture-density head to a posterior with
    shaded tail mass, then posterior-sized positions; a dashed
    branch shows training on simulator draws.}  \caption{The Neural Filter Estimator: the panel with its
    cross-sectional and time-series steps (left); the posterior head
    and the pre-specified decision layer (right); dashed, training on
    simulated draws and the conformal layer.}
  \label{fig:pipeline}
\end{figure*}

\subsection{Notation}
\label{sec:method:notation}

We consider $N$ assets observed on trading days $t = 1, \dots, T$,
with daily returns $R_t \in \mathbb{R}^N$ and $M$ characteristics
$X_t \in \mathbb{R}^{N \times M}$, following the notation of
\citet{epstein-2025-attentionfactors}. A characteristic is an
attribute of a firm known at the start of day $t$, such as its
market capitalization, book-to-price ratio or recent return and
volatility. From returns and characteristics a
residual return $\varepsilon_{i,t}$ is constructed for each asset,
and its cumulative sum $S_{i,t} = \sum_{s \le t} \varepsilon_{i,s}$
is the spread. We drop the asset index $i$ where it is immaterial.
Two horizons are used throughout: the filter reads a look-back
window of $L$ trading days, and a position is held for $H$ trading
days ($L = 60$ and $H = 20$ in Section~\ref{sec:experiment}).

\subsection{Data-Generating Process}
\label{sec:method:model}

The filter is trained by neural posterior estimation
(Section~\ref{sec:method:training}): on simulated draws in which the
latent state is known, it learns the map from an observed window to
the posterior over the state, and it is then applied to real windows
without further fitting. This amortization requires a process to
simulate from, and the posterior is worth only as much as that
process resembles the data, so the data-gen\-er\-at\-ing process is
the object on which the method rests and must be stated explicitly.
Conventional strategies assume one implicitly when they expect the
spread to revert to a stationary level, and the end-to-end
approaches leave theirs in the training distribution. 

The process describes the increment of the spread: since the spread
is a cumulative sum of residual returns, its daily increment is the
residual return itself, and mean reversion means that this increment
tends to move the spread back toward a level $m_{i,t}$. We write the
law of the increment conditional on where the spread stands, indexed
by a latent variable $z_{i,t} \in \mathcal{Z}$ and by parameters
$\theta$,
\begin{equation}
S_{i,t+1} - S_{i,t} \;\sim\;
  p\bigl(\,\cdot \mid S_{i,t};\, \theta,\, z_{i,t}\bigr).
  \label{eq:kernel}
\end{equation}
Given the latent variables, the spread is Markov: the next increment
depends on the past only through $S_{i,t}$, so a filter can carry its
state forward one day at a time, and $z_{i,t}$ selects which
conditional law is in force on day $t$. The following definition
fixes which values of the index count as regimes.

\begin{definition}[Regime]\label{def:regime}
Let the increment of a spread follow \eqref{eq:kernel} with latent
index $z_{i,t} \in \mathcal{Z}$, and for a horizon $H$ let
\begin{equation}
\rho_H(z) \;=\;
  \mathbb{E}\!\left[\frac{S_{i,t} - S_{i,t+H}}{S_{i,t} - m_{i,t}}  \,\middle|\, S_{i,t},\, m_{i,t},\, z_{i,t} = z\right]
  \label{eq:rho}
\end{equation}
be the expected fraction of the deviation from the level that is
reversed over the next $H$ days when the index equals $z$. Given a
threshold $\rho^\ast \in (0,1)$, the \emph{regimes} of the process
are the two classes $\{z : \rho_H(z) > \rho^\ast\}$ and
$\{z : \rho_H(z) \le \rho^\ast\}$ of $\mathcal{Z}$.
\end{definition}

Different regimes thus have different conditional laws, as in
\citet{hamilton-1989-regime}, and a change of index is a change of
regime only if it changes what a trade can expect. We instantiate the kernel as
Ornstein--Uhlenbeck (OU) mean reversion with regime-indexed speed,
\begin{equation}
S_{i,t+1} - S_{i,t}
  \;=\; \kappa(z_{i,t})\,\bigl(m_{i,t} - S_{i,t}\bigr)
  \;+\; \sigma_{i,t}\,\zeta_{i,t},
  \label{eq:sou}
\end{equation}
where the innovation scale $\sigma_{i,t}$, the level $m_{i,t}$ and
the law of the innovation $\zeta_{i,t}$ are themselves latent
processes (Appendix~\ref{app:dgp}). For $\kappa$ constant over the
horizon, $\rho_H = 1 - e^{-\kappa H}$ (exact in continuous time), so
a threshold on $\rho_H$ is a threshold $\tau^\ast$ on the half-life
$\tau = \ln 2 / \kappa$, and the two regimes are $\textup{C} =
\{\tau \le \tau^\ast\}$ and $\textup{D} = \{\tau > \tau^\ast\}$,
converging and diverging. We set $\tau^\ast = L = 60$ days, because
a slower reversion cannot be told from none within the window and
the decision cannot act on it; with $H = 20$ this is $\rho^\ast = 1 -
2^{-H/\tau^\ast} \approx 0.21$, so in C at least a fifth of the
deviation reverts within the holding period and in D less, down to
the boundary $\kappa = 0$ at which the pull vanishes. The two speeds
$\theta = (\kappa(\textup{C}), \kappa(\textup{D}))$ lie on either
side of $\tau^\ast$ (half-lives of $10$ and $120$ days in the
simulations of Section~\ref{sec:experiment}); the hyper-parameters of
the innovation processes are fixed by moment matching
(Appendix~\ref{app:dgp}). The regime path follows a first-order chain with transition matrix
$A$, so the prior factorizes as $p(\theta, z_{1:T}) = p(\theta)\,
p(z_1) \prod_{t=2}^{T} A_{z_{t-1} z_t}$.

The mean-reversion kernel is the standard model of the literature
this work extends \cite{avellaneda-2010-statarb}, and adopting it
entails little loss of generality: the filter can be trained on any
kernel of \eqref{eq:kernel} that can be simulated, and the direct
variant of Section~\ref{sec:method:nfe} drops the kernel's link
between the state and the forecast. The innovation law, by contrast, is chosen by a two-sample test
\cite{lopez-paz-2017-c2st}: a candidate law is retained only if a
classifier cannot separate its simulated windows from real residual
windows beyond a pre-specified threshold, and only stochastic
volatility with asymmetric innovations passes on both
cross-sections (Appendix~\ref{app:dgp}).

The regime is a latent variable of the process; the data provide
only an estimate of it. With the whole path in
hand the best estimate is the smoothed assignment $\hat{z}_{i,t} =
\arg\max_{z} p(z_{i,t} = z \mid S_{i,1:T}, \hat{\theta})$,
available only ex post. What a trader needs at time $t$ is the
\emph{filtering} posterior, which conditions only on the past,
\begin{equation}
p(\theta, z_t \mid S_{1:t})
  \;=\; \frac{p(S_{1:t} \mid \theta, z_t)\, p(\theta, z_t)}
             {\sum_{z'} \int p(S_{1:t} \mid \theta', z')\,
              p(\theta', z')\, \mathrm{d}\theta'}.
  \label{eq:posterior}
\end{equation}
Its likelihood is marginalized over the earlier states and the
latent scale and level paths, so the posterior is intractable;
computing it is the filtering problem of state-space models. The NFE
is a neural network trained to return the marginal of this posterior
that the decision uses, the posterior over the current speed, in a
single forward pass for every window and every asset: an amortized
filter.

\subsection{Neural Filter Estimator}
\label{sec:method:nfe}

Once the process is fixed, the question is what information to give
the filter at time $t$. A single spread's window identifies its
state only partly, since the level must be estimated from the same
window; the rest of the panel, the same day's returns and
characteristics of the other assets, enters through the
cross-sectional step. As shown in Figure~\ref{fig:pipeline},
the panel is a table with days as rows and assets as columns, and
the NFE processes it along both axes: a cross-sectional step across
the assets of one day, which removes what the assets share and
leaves each asset's own spread, and a time-series step along each
asset's window, which reads the spread's dynamics. The tabular foundation models \cite{qu-2025-tabicl, kong-2026-tabfm}
organize in-context learning on tables the same way, and the
cross-sectional step corresponds to their row embedding
(Section~\ref{sec:related}), with one deliberate difference: those
models treat rows as exchangeable, which a time-ordered panel is
not, so the NFE's time-series step is a recursion that keeps the
order of days, while its cross-sectional step, a factor model, is
indifferent to the order of assets.

\emph{Cross-sectional step.} The classical construction uses returns
alone: the factors are the leading principal components of the
recent return covariance, and each asset's residual is what they
leave. The Attention Factor Model (AFM)
\cite{epstein-2025-attentionfactors} instead computes the factor
weights each day by attention over the characteristics, so that
assets with similar characteristics are pooled into the same factor
portfolio and the factors adapt as the characteristics change. In
either construction a factor model maps returns and characteristics
to one residual return per asset,
\begin{equation}
F_t = \Omega_{t-1} R_t, \qquad
\varepsilon_t = R_t - \beta_{t-1}^{\top} F_t,
  \label{eq:rowstep}
\end{equation}
where $\Omega_{t-1}$ holds the factor weights and $\beta_{t-1}$ the
loadings, both estimated with information up to $t-1$; the
cumulative sum of $\varepsilon_t$ is the spread the time-series step
reads. The filter takes this residual from whichever factor model
supplies it rather than learning its own: the factor construction is
where the earlier generations of the lineage in
Section~\ref{sec:related} differ, and holding it fixed isolates
the contribution of the posterior. Section~\ref{sec:experiment}
evaluates both constructions; the NFE of that section uses AFM's factors, refit
annually as in the AFM strategy.

\emph{Time-series step.} A filter is a recursion: the estimate of
the state is updated from the previous estimate and each new
observation without revisiting the past, which is what makes it
causal. A recurrent network has this form, so we encode each
spread's look-back window with a gated recurrent unit (GRU), $h_j =
\operatorname{GRU}_\psi(h_{j-1}, w_{i,j})$ for $j = t-L+1, \dots,
t$, where $w_{i,j}$ is the day-$j$ entry of the window $w_{i,t}$,
the spread and its increment standardized by the window's own mean
and standard deviation, and the final state $h_t$ summarizes the
window.
The standardization makes the input invariant to location and
scale, so the filter reads the shape of the window and not its
units, and the simulator has to match the data in shape only. A panel form,
both steps as alternating attention in one network, is tested only
on simulated draws in which the latent truth is known (truth-known
draws hereafter) and is left for future work
(Section~\ref{sec:experiment:ident}).

\emph{Posterior head.} The state the decision needs is the half-life
of the current regime, $\tau_t = \ln 2 / \kappa(z_t)$. From the
final state $h_t$ a mixture-density head outputs a posterior
$q_\psi(\tau \mid w_{i,t})$, a five-component Gaussian mixture in
$\log_2 \tau$ whose weights, locations and widths are functions of
$h_t$. The half-life spans two orders of magnitude, hence the
logarithmic scale, and a mixture rather than a single Gaussian lets
an ambiguous window spread its mass over both regimes instead of
committing to one. The regime probability of Definition~\ref{def:regime} is the
tail mass of this posterior beyond $\tau^\ast$,
\begin{equation}
P\bigl(z_{i,t} = \textup{D} \mid w_{i,t}\bigr)
  \;=\; \int_{\tau^\ast}^{\infty}
        q_\psi\bigl(\tau \mid w_{i,t}\bigr)\, \mathrm{d}\tau,
  \label{eq:head}
\end{equation}
so the regime probability and the half-life posterior are consistent
by construction. The decision layer uses the half-life posterior through its
moments; the regime probability is reported but is not an input to
the sizing rule. A \emph{direct variant} places the same head on the
standardized $H$-day increment of the spread, $q_\psi(\Delta S \mid
w_{i,t})$. It is a control for the functional form and not a
deployable filter, since the decision layer needs a posterior over
the half-life: the OU-form posterior ties the predictive mean to the
half-life through the kernel \eqref{eq:sou}, and the comparison of
Section~\ref{sec:experiment:ident} tells whether that form limits
the forecast.

\subsection{Decision layer}
\label{sec:method:decision}

The decision layer is pre-specified: the rules below were fixed
before any run and apply with the same conventions to every
strategy, and the filter is never fitted to realized profit, so the
posterior is the only interface between the filter and the
portfolio. If the rule were learned, as in AFM's end-to-end trading
network, a difference in performance could come from the rule as
much as from the state estimate; with the rule fixed, the strategies
differ only in the residual they trade and in whether the position
is sized from a point estimate or from the posterior. Three
components are selected on the training split: the threshold rule's
entry level and AFM's characteristic set, by their training-split
Sharpe ratios, and the per-asset prior below, from the training
split's half-lives.

The signal is the standardized deviation $d$, the last entry of the
window's spread channel: the spread's distance from its window mean
in units of the window's standard deviation, clipped at $\pm 3$.
This is the signal of the classical pipeline
\cite{avellaneda-2010-statarb}. The base rule sets the position
proportional to $-d$, a contrarian position against the deviation
with no tuned parameter; the threshold rule of
\citet{avellaneda-2010-statarb} opens a unit position once $|d|$
exceeds an entry threshold tuned on the training split. Both rules act on the deviation as if it would revert; the NFE
instead sizes positions on the same signal from the posterior. Under the kernel \eqref{eq:sou} with reversion rate
$\kappa$, the expected change of the spread over the holding period
is $-d\,(1-e^{-\kappa H})$ in standardized units and, with the scale
and level at their window estimates, its variance is
$1-e^{-2\kappa H}$; when $\kappa$ is uncertain, the law of total
variance adds the variance of that expected change across the
posterior, and the mean--variance optimal position is the expected
change divided by the total variance,
\begin{equation}
\omega_{i,t} \;\propto\;
  \frac{-\,d\;\mathbb{E}\!\left[1-e^{-\kappa H}\right]}
       {\mathbb{E}\!\left[1-e^{-2\kappa H}\right]
        + d^{2}\operatorname{Var}\!\left(e^{-\kappa H}\right)},
  \label{eq:sizing}
\end{equation}
with all moments taken under $q_\psi(\cdot \mid w_{i,t})$ through
$\kappa = \ln 2 / \tau$. The term
$d^{2}\operatorname{Var}(e^{-\kappa H})$ is identically zero for any
point estimate, so \eqref{eq:sizing} with a point estimate is the
plug-in rule of the classical pipeline; Section~\ref{sec:experiment}
compares the posterior with the base rule on the same residuals and,
at matched signal and window, with this plug-in rule. The term
shrinks the position when the posterior is wide and leaves it
unchanged when the posterior is narrow. Positions are then demeaned
across assets and scaled to unit gross exposure, the dollar-neutral
convention under which every strategy is evaluated, so within a
fixed gross the penalty moves exposure from spreads whose posterior
is wide to those whose posterior is narrow. It accounts for
uncertainty in the reversion speed only; the level appears in $d$ as
a window estimate and is not penalized, the omission
Section~\ref{sec:experiment:ident} finds to be the largest source of
identification loss.

The training prior over $\tau$ is deliberately wide, flat in $\log_2
\tau$ over $[3, 400]$ days, so that one network serves every spread. At deployment it is
replaced, for each stock, by a narrower Gaussian prior in $\log_2
\tau$ centred on that stock's own half-life, estimated on $L$-day
blocks of the training split and shrunk toward the panel mean by
empirical Bayes, so that a stock with few or noisy blocks is pulled
further toward the mean. Because the head is trained under a flat
prior, its output is likelihood-shaped, and the new prior is applied
by reweighting the posterior draws used in \eqref{eq:sizing}; the
network is not retrained.

\subsection{Training}
\label{sec:method:training}

Real data never reveal the latent state, but simulated draws include
it by construction, so the simulator supplies the labels. The filter
is trained as in neural posterior estimation
\cite{papamakarios-2016-npe}, by maximizing the likelihood of the
known half-life under the head, $\mathcal{L}(\psi) =
-\,\mathbb{E}_{(\tau, w) \sim p_{\textup{train}}} [\log
q_\psi(\tau \mid w)]$, on windows drawn from the process of Section~\ref{sec:method:model}
with one simplification: in training the half-life is held fixed
over the window and drawn uniformly in $\log_2 \tau$ on $[3, 400]$
days, with the innovation hyper-parameters at their moment-matched
values (Table~\ref{tab:setup}). Each window then carries one half-life, so $\theta$ reduces to
$\tau$, and the flat prior gives the likelihood-shaped head on which
the per-asset prior relies. The
two-state chain of Section~\ref{sec:method:model} is the process of
the two-sample test and of the identification experiments of
Section~\ref{sec:experiment:ident}; the calibration check of
Figure~\ref{fig:cumret}(b) uses the training draws themselves. No real returns enter the
training of the network. This keeps the posterior checkable: its calibration is measured on
simulated draws, where the truth is known
(Section~\ref{sec:experiment:ident}), and it cannot have been fitted
to the outcomes it is later scored on. Real data enter the filter only through the choice of innovation
law, by the two-sample test on full-sample windows
(Appendix~\ref{app:dgp}), and the moment matching of its
hyper-parameters on the training split; at deployment they enter
through the per-asset prior, the split-conformal layer
\cite{romano-2019-cqr} on every learned predictive distribution of
Section~\ref{sec:experiment:ident}, and the direct variant's target
scale, all fitted on the training split
(Appendix~\ref{app:details}).

\section{Experiment}
\label{sec:experiment}

We report one empirical study: walk-forward statistical-arbitrage
portfolios on $120$ Korea Exchange (KRX) stocks over $2010$--$2026$,
with annual refits (Table~\ref{tab:setup}), net of transaction and
shorting costs; the package, the experiment scripts and the data
pipeline are available at \url{https://github.com/minkeymouse/nfe}.
One strategy represents each
generation of the line of work in Section~\ref{sec:related}. Fama--French
three-factor characteristic residuals (FF3) \cite{fama-french-1993},
rolling principal-component residuals (PCA) and Attention Factor
Model residuals (AFM) \cite{epstein-2025-attentionfactors} are each
traded by the same base rule; the NFE, posterior sizing on the AFM
factors, is the fourth generation. In Table~\ref{tab:book}, AFM denotes AFM's residuals under the base
rule in our own implementation; its learned trading network is
reported among the variants. We compare orderings, not levels, so
absolute Sharpe ratios are not compared with the published U.S.\
figures. The conventions, fixed in advance and identical across strategies,
are collected in Table~\ref{tab:setup} and
Appendix~\ref{app:details}, and every label-free diagnostic is
reported with its null model and positive control. Results come from
one training seed and are exploratory unless stated; a result is
called \emph{confirmed} when it was reproduced on a disjoint period
or on disjoint seeds against criteria fixed beforehand.

\subsection{Results}
\label{sec:experiment:trading}

Table~\ref{tab:book} reports the later period, $2018$ to $2026$,
on the days after its training split, one row per generation, with
a market-adjusted residual as a negative control; $\Delta$SR is the
annualized Sharpe ratio of the daily return difference to the random
benchmark (Appendix~\ref{app:details}).

\begin{table}[t]  \caption{Out-of-sample performance, KRX panel, 2018-04 to 2026-07
    (evaluation days), net of costs; bold marks the best of the four
    generations.}
  \label{tab:book}
  \footnotesize
  \setlength{\tabcolsep}{2.6pt}
  \begin{tabular}{l rrrrrr}
    \toprule
    strategy & SR & $\mu$ (\%) & $\sigma$ (\%) & MDD (\%) & turnover & $\Delta$SR \\
    \midrule
    FF3 & $-0.12$ & $-1.1$ & $8.8$ & $23.0$ & $\mathbf{0.052}$ & $-0.01$ \\
    PCA & $0.57$ & $3.8$ & $\mathbf{6.7}$ & $7.1$ & $0.053$ & $+0.59$ \\
    AFM & $1.45$ & $10.4$ & $7.2$ & $4.2$
      & $0.055$ & $+1.35$ \\
    NFE & $\mathbf{1.49}$ & $\mathbf{10.6}$ & $7.1$ & $\mathbf{4.1}$
      & $0.055$ & $\mathbf{+1.38}$ \\
    \midrule
    market-adjusted & $-0.12$ & $-1.2$ & $9.2$ & $21.8$ & $0.052$
      & $-0.01$ \\
    random benchmark & $-0.23$ & $-1.0$ & $4.4$ & $16.0$ & $0.071$
      & \\
    \bottomrule
  \end{tabular}
\end{table}

\emph{The four generations.} FF3's fixed characteristics earn no excess return over the random
benchmark, and neither does the market-adjusted control.
Statistical factors rank above the benchmark (PCA, Sharpe $0.57$) and
learned factors higher still (AFM, $1.45$). In the earlier period no strategy outperforms the
benchmark. The
ordering therefore depends on the period, and because the panel is
selected at the end of the sample, the levels are an upper bound.

\begin{figure}[t]
  \centering
  \begin{minipage}[b]{0.56\columnwidth}
    \centering
    \includegraphics[width=\linewidth]{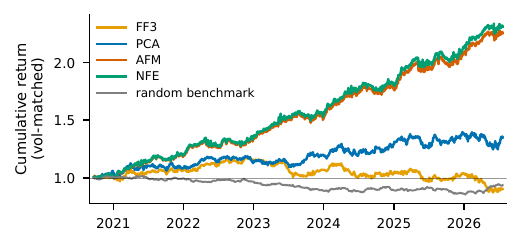}\\[-2pt]
    {\footnotesize (a)}
  \end{minipage}\hfill
  \begin{minipage}[b]{0.42\columnwidth}
    \centering
    \includegraphics[width=\linewidth]{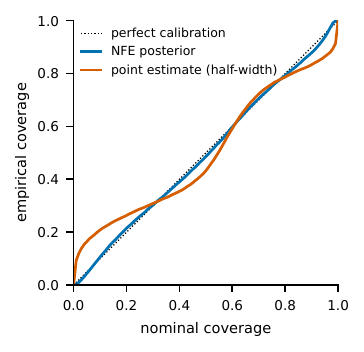}\\[-2pt]
    {\footnotesize (b)}
  \end{minipage}
  \Description{Left: volatility-matched cumulative return of the four
    generations and the random benchmark over 2018-04 to 2026-07.
    Right: reliability diagram of empirical against nominal coverage
    on truth-known draws; the NFE posterior tracks the diagonal and
    the same posterior at half its width bows below it.}
  \caption{(a) Cumulative net return of the strategies of
    Table~\ref{tab:book}, scaled to a common volatility; the random
    benchmark at its own scale. (b) Coverage of the true half-life on
    truth-known draws: the NFE posterior and the same posterior at
    half its width.}
  \label{fig:cumret}
\end{figure}

\emph{The fourth generation.} The NFE matches AFM's Sharpe, a
difference in level of $+0.04$, within the period's noise, at slightly lower volatility and drawdown and with equal turnover.
The standard error of either strategy's $\Delta$SR is $0.41$, so a
difference of this size is not resolvable here. Posterior sizing
also raises the PCA-residual strategy's Sharpe ratio from $0.57$ to
$0.68$, a difference subject to the same noise. The per-asset prior of Section~\ref{sec:method:decision} was
confirmed on three disjoint training seeds. With the training prior
the NFE's mean Sharpe ratio is $1.424$ in this period and $-0.259$
in the earlier one; with the per-asset prior it is $1.470$ and
$-0.193$; with the per-asset centres permuted across stocks it is
$1.458$ and $-0.215$. The gains are consistent in sign across seeds but small relative to
the period's noise, and most of them come from calibrating the
prior's centre and width to the panel rather than from the per-stock
centres. Figure~\ref{fig:cumret}(a) shows the cumulative returns.

\emph{Window length.} A posterior over the reversion speed
conditioned on a $250$-day window, applied to the same $60$-day
signal, has a larger $\Delta$SR than AFM on two training seeds. The gain over AFM,
however, is a window effect: a pre-specified control that trades the
$250$-day deviation directly reaches a Sharpe ratio of $2.77$ in
this period, above every strategy, and $-0.91$ in the other, so we
do not claim it and retain the $60$-day window. What survives at
matched signal and window is that this posterior outperforms the
plug-in rule with a $250$-day least-squares point estimate of the
speed.

\emph{Other variants.} A cross-sectional step learned on the panel
itself, asset attention trained within the NFE in place of AFM's
factors, reaches a Sharpe ratio of $-0.03$, within the random
benchmark's noise, so the NFE keeps AFM's factor extraction. Among
the rule variants, the threshold rule reaches $0.75$ on PCA
residuals, above both the base rule and posterior sizing there, and
AFM's own end-to-end trading network, evaluated under this paper's
$60$-day window and $20$-day holds instead of its native daily
rebalancing, has the lowest Sharpe ratio of any strategy on its
factors ($0.25$); under a convention that is not its own, that row
is not a replication of \citet{epstein-2025-attentionfactors}, and
the NFE adopts AFM's factors and not its trading network.

\emph{Overshoot.} One secondary finding contradicts the pre-stated
prediction that the contrarian rule earns more after a window
labeled converging than after one labeled diverging: with the label
computed from the deviation about a level $g(X)$ learned from
characteristics (Appendix~\ref{app:dgp}; this pre-registered label
is not the posterior of \eqref{eq:head}), the base rule's realized
P\&L per holding period is $+118$\,bp after a diverging window and
$+27$\,bp after a converging one, a converging-minus-diverging
difference of $-92 \pm 25$\,bp. The difference survives its
controls: the injected positive control recovers the predicted sign
($+81 \pm 20$), the random-walk null shows no difference ($+8 \pm
21$), the ordering persists when P\&L is normalized by position
weight, which rules out a sizing artifact, and it reproduces on a
disjoint period ($-90 \pm 37$, with the fundamentals at their
neutral fill before $2016$). On this panel a $60$-day divergence
behaves as an overshoot: the label orders expected return within a
profitable rule and does not separate a profitable state from an
unprofitable one, so its use is as a sizing input and not as a
signal to stand aside. The sizing rule \eqref{eq:sizing} does not use it, and the
shrinkage it applies to slow-reversion windows runs against this
finding, which is left open.

\subsection{Analysis}
\label{sec:experiment:ident}

\emph{Identification.} On truth-known draws of the special case of
the generative model with Gaussian innovations of constant scale,
the only case in which the exact likelihood-ratio detector can be
computed, every practical estimator and the reference detectors fall
within one band of $0.61$--$0.66$ AUC. The
Bayes bound, i.e.\ the exact likelihood-ratio detector given the
true level and scale, computed on draws whose state is constant over
the window, reaches $0.93$; any detector that sees only the window
is bounded by it. In accuracy the amortized posterior is indistinguishable from the
plug-in estimator of \citet{avellaneda-2010-statarb}: the window
limits accuracy whatever the estimator, and the largest loss comes
from the unknown level, since with the level known separation rises
from $0.62$--$0.64$ to $0.79$--$0.81$. Over $60$ days the slow
state (D, half-life $120$ days) completes $29\%$ of its reversion
and the fast state (C, $10$ days) $98\%$, so a level estimated from
the window sits near the current value and makes the slow state
resemble the fast one. In a separate input-channel experiment, a channel carrying the
un-demeaned spread closes almost none of that gap: over three seeds
it adds $+0.0002 \pm 0.0005$ AUC alone and $+0.0010 \pm 0.0004$
together with the demeaned channel, against $+0.139 \pm 0.002$ when
the encoder is given the true level. On the half-life itself, an unamortized particle filter over the
same model ($25 \times 256$ particles) is the more accurate, with
rank correlation $0.74$ against $0.59$ for the NFE on truth-known
draws, at $7.9$\,s against $2.5$\,ms per cross-section on a desktop
CPU (training the NFE once takes $104$\,s); the NFE trades that
fidelity for amortization.

\emph{Calibration.} The paper's central claim is that the posterior
is calibrated. Figure~\ref{fig:cumret}(b) tests the half-life posterior
on truth-known draws under the training prior, before the per-asset
reweighting, which is not separately checked: its coverage tracks
the nominal level throughout, so the regime probability
\eqref{eq:head}, the mass of the same posterior beyond $\tau^\ast$,
is calibrated in the same marginal sense. The same posterior at half
its width under-covers at every level: the maximum deviation from
nominal is $0.022$ on the seed shown and at most $0.010$ on two
further seeds, against $0.08$--$0.12$ at half width. A posterior that ignored the window would also pass this check, so
it is read together with the identification band above. On real
data, the predictive distribution of the $H$-day change is scored on
the held-out rolling-PCA residual series, which need no learned
factors, with identical features and the same conformal layer for
every method, against an unconditional baseline
$\mathcal{N}(0,\hat\sigma\sqrt{H})$, $\hat\sigma$ the asset's trailing
$60$-day volatility.
Table~\ref{tab:honesty} reports the continuous ranked probability
skill score against the baseline (CRPSS), the dispersion of the
predictive mean relative to the variance of the outcome, and the
coverage and width of the nominal $90\%$ interval, crediting
sharpness only subject to calibration \cite{gneiting-2007-sharpness}.
Its last column is the \emph{forecast--realized gap}: with positions
formed for every row by the same mean--variance rule from that row's
predictive mean and variance, the strategy's own forecast of its
return over the holding period minus the return it realizes, in
basis points per holding period at unit gross. This is the
postdecision surprise of \citet{smith-2006-optimizer}, the
optimizer's curse, and the decision-level form of the winner's curse
of Section~\ref{sec:introduction}: a forecast that is too confident
about the positions it selects overstates what they earn. The gap is
smallest for a forecast of zero, so it is read together with the
forecast's dispersion.

\begin{table}[t]  \caption{Predictive distributions of the $H$-day change on held-out
    KRX PCA residuals, 2018-04 to 2026-07 (evaluation days).}
  \label{tab:honesty}
  \footnotesize
  \setlength{\tabcolsep}{2.4pt}
  \begin{tabular}{l rrr rr}
    \toprule
    predictive distribution & CRPSS & disp. & cov.\ & width & gap (bp) \\
    \midrule
    unconditional baseline & $0.00$ & $0.00$ & $0.89$ & $2.30$ & $0$ \\
    OU point estimate & $-0.97$ & $0.75$ & --- & --- & $1084$ \\
    mixture density net & $-0.50$ & $0.71$ & $0.87$ & $3.99$ & $798$ \\
    amortized OU-form (ours) & $-0.11$ & $0.22$ & $0.89$ & $2.40$ & $588$ \\
    amortized direct (ours) & $0.00$ & $0.02$ & $0.89$ & $2.17$ & $137$ \\
    boosted quantiles & $-0.03$ & $0.08$ & $0.88$ & $2.28$ & $277$ \\
    \bottomrule
  \end{tabular}
\end{table}

\begin{figure}[t]
  \centering
  \includegraphics[width=0.8\columnwidth]{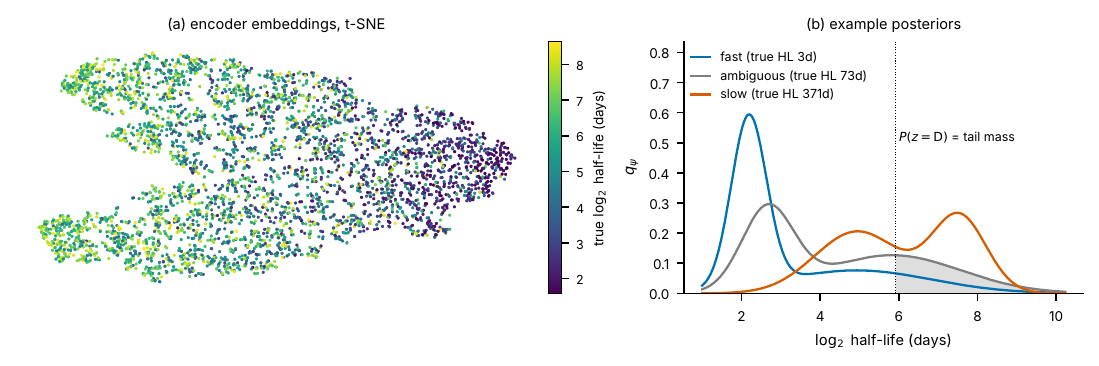}
  \Description{Left: t-SNE scatter of encoder embeddings colored by
    true half-life. Right: three example posterior densities with the
    slow-state tail shaded.}  \caption{(a) t-SNE of encoder states on truth-known draws, colored
    by the true $\log_2$ half-life. (b) Posteriors for a fast, an
    ambiguous and a slow window; dotted, $\tau^\ast = 60$ days.}
  \label{fig:posterior}
\end{figure}

On real data, as reported in Table~\ref{tab:honesty}, no method has
positive skill against the baseline, and the point estimate and the
mixture density net score well below it, with returns alone or
with fundamentals added (the latter not tabulated). Where the methods differ is in how accurately they assess their own
error. The OU point estimate, the unshrunk window plug-in that
mean--variance sizing requires, has the largest forecast dispersion
in the table and gives no interval; its forecast exceeds
its realized return by roughly a thousand basis points per holding
period. The gap of the amortized methods is roughly two to eight
times smaller, and the ordering holds on three training seeds:
$572$--$593$\,bp for the OU-form posterior and $120$--$150$ for the
direct variant, against $1084$ for the point estimate and
$658$--$838$ for the mixture density net. The reduction is a uniform
shrinkage rather than a conditional one: the gap falls with forecast dispersion in every row of
Table~\ref{tab:honesty}, and a point forecast scaled by a single
constant, chosen so that its mean absolute forecast on the training
split equals the posterior's, reaches the posterior's gap within
sampling error. The OU-form posterior's gap minus the rescaled
point forecast's is $-1$ to $-5$\,bp over three seeds on truth-known
draws of the admitted process, and $+1 \pm 17$ and $-13 \pm 12$\,bp
in the two periods of a KRX panel rebuilt under the same universe
rule from a second price source; on the no-reversion null it is
$-38 \pm 10$ and $-42 \pm 10$\,bp, so there the posterior is the
more cautious (Appendix~\ref{app:details}). The gap that remains for every method we attribute to the level
estimated from the window, the loss identified by the level-known comparison above. The conformal
layer brings every learned method to nominal coverage, but coverage
says nothing about the conditional mean. The amortized OU-form posterior is constrained by its functional
form: its forecast dispersion on the volatility-matched random-walk
null (Table~\ref{tab:setup}) is as large as on real data. The
amortized direct variant matches the baseline's skill with a
narrower interval, has the smallest gap of any learned method, and shows no dispersion on that null. Figure~\ref{fig:posterior} shows the mechanism on
truth-known draws: the encoder state is organized by the latent
half-life, which it never receives as input, and an ambiguous window
yields a wide posterior and, through \eqref{eq:sizing}, a smaller
position; whether the sizing results of
Section~\ref{sec:experiment:trading} come from the width varying
across windows or from the dependence of the penalty on $d^{2}$,
which reweights large deviations even at a fixed width, has not been
separated on real data.

\emph{Panel form.} On truth-known panel draws, the panel form of
Section~\ref{sec:method:nfe}, which is not the filter used in the
portfolio, outperforms the recurrent time-series step, and by as
much on panels whose innovation hyper-parameters are independent
across assets, so that margin reflects capacity and not pooling.
Pooling is measured separately, as the contrast between panels whose
levels share a common factor and an independent-factor control. On
shared-factor panels the held-out negative log-likelihood of the
half-life falls from $2.088$ to $1.762$, of which $0.179 \pm 0.003$
is the contrast with the independent-factor control; the panel form
recovers most of that factor and stays calibrated, and the true
innovation hyper-parameters, the most that pooling a shared
volatility state could supply, bring no further improvement, so the
gain comes from the shared factor. The contrast reproduces on
disjoint seeds ($0.170 \pm 0.002$) in a confirmation run that also
replaces the exploration's rolling-PCA cross-sectional step by asset
attention.

\section{Conclusion}
\label{sec:conclusion}

This paper nowcasts the state of a residual spread instead of
forecasting its next move. It builds on the classical and the
end-to-end approaches to statistical arbitrage, trains its filter on
a simulated process whose innovation law was chosen by testing
candidates against real residuals, and uses the uncertainty of the
estimate in the trading decision. Three
measured results follow. Within them, four findings have passed a confirmation
run: the overshoot finding, the panel-form pooling contrast, the per-asset prior and the ordering of the forecast--realized gaps. 

First, the regime is identifiable, but only marginally: every
practical estimator and the strong reference detectors fall in one
narrow band near $0.61$--$0.66$ AUC, against $0.93$ for a fully
informed detector, and the largest loss comes from estimating the
level from the same window as the reversion. Second, a calibrated
posterior gives an accurate account of its own error. A point
estimate's forecast exceeds its realized return by roughly a thousand
basis points per hold and gives no interval; the amortized methods
reduce that gap by a factor of roughly two to eight, and since a point
forecast rescaled by one constant reaches the same gap, the reduction
is a uniform shrinkage, and the interval, which the point estimate
lacks, is what the posterior adds.
No method shows skill
against an unconditional baseline, with returns alone or with
fundamentals added. Third, in the portfolio evaluation the residual
construction matters more than the trading policy. Posterior sizing
matches AFM's Sharpe, within the period's noise, at no more risk. No
pre-specified decision-layer configuration adds return of the size
the residual construction adds; the gain from a longer
window is unstable across periods and not claimed; and the fully
learned
cross-sectional step does no better than the random benchmark.

One methodological lesson runs through all three results: a
label-free diagnostic needs a null model and a positive control
before its output can be interpreted. Without them several of ours
lacked power or were biased upward, and we withdrew the conclusions
drawn from them, the $250$-day gain among them. A simulator can pass a two-sample test that has almost no power
against the kernel it appears to validate: changing the kernel moves
the AUC by at most $0.014$ (Appendix~\ref{app:dgp}).

Since the largest identification loss comes from the level estimated
from the same window, the natural next step is the panel form of the
NFE: one causal network over the cross-section and time, pretrained
on a panel-level generative model, that pools the cross-section
inside the network and returns its posteriors jointly, with a
selective state-space encoder \cite{gu-2023-mamba} as its
time-series step. On truth-known draws with a shared factor in the
levels, the panel form already improves the half-life posterior over
the per-asset filter, in part through capacity, and stays calibrated
(Section~\ref{sec:experiment:ident}); on real data the two
cross-sectional routes tried here, a learned cross-sectional step and
a level learned from characteristics, did not help, so its value
there is untested. Whether on real data the
information is worth more than the cost of acting on it is the
question this work leaves open.

\begin{acks}
We thank Jae Young Kim (Seoul National University) and Young Jae Cho
(ATZ Trading) for their guidance and support, and the three anonymous
ICAIF reviewers for their careful reading and constructive comments.
\end{acks}

\section*{Ethics and Privacy Statement}
This work uses only public daily market data and simulated data; no
vendor prices are redistributed and no personal data is used.

\bibliographystyle{ACM-Reference-Format}
\bibliography{reference}


\begin{thebibliography}{27}

%%% the Web address.

\ifx \showCODEN    \undefined \def \showCODEN     #1{\unskip}     \fi
\ifx \showISBNx    \undefined \def \showISBNx     #1{\unskip}     \fi
\ifx \showISBNxiii \undefined \def \showISBNxiii  #1{\unskip}     \fi
\ifx \showISSN     \undefined \def \showISSN      #1{\unskip}     \fi
\ifx \showLCCN     \undefined \def \showLCCN      #1{\unskip}     \fi
\ifx \shownote     \undefined \def \shownote      #1{#1}          \fi
\ifx \showarticletitle \undefined \def \showarticletitle #1{#1}   \fi
\ifx \showURL      \undefined \def \showURL       {\relax}        \fi
% invisible to TeX
\providecommand\bibfield[2]{#2}
\providecommand\bibinfo[2]{#2}
\providecommand\natexlab[1]{#1}
\providecommand\showeprint[2][]{arXiv:#2}

\bibitem[Avellaneda and Lee(2010)]%
        {avellaneda-2010-statarb}
\bibfield{author}{\bibinfo{person}{Marco Avellaneda} {and}
  \bibinfo{person}{Jeong-Hyun Lee}.} \bibinfo{year}{2010}\natexlab{}.
\newblock \showarticletitle{Statistical Arbitrage in the {US} Equities Market}.
\newblock \bibinfo{journal}{\emph{Quantitative Finance}} \bibinfo{volume}{10},
  \bibinfo{number}{7} (\bibinfo{year}{2010}), \bibinfo{pages}{761--782}.
\newblock
\href{https://doi.org/10.1080/14697680903124632}{doi:\nolinkurl{10.1080/14697680903124632}}


\bibitem[Cranmer et~al\mbox{.}(2020)]%
        {cranmer-2020-sbi}
\bibfield{author}{\bibinfo{person}{Kyle Cranmer}, \bibinfo{person}{Johann
  Brehmer}, {and} \bibinfo{person}{Gilles Louppe}.}
  \bibinfo{year}{2020}\natexlab{}.
\newblock \showarticletitle{The Frontier of Simulation-Based Inference}.
\newblock \bibinfo{journal}{\emph{Proceedings of the National Academy of
  Sciences}} \bibinfo{volume}{117}, \bibinfo{number}{48}
  (\bibinfo{year}{2020}), \bibinfo{pages}{30055--30062}.
\newblock
\href{https://doi.org/10.1073/pnas.1912789117}{doi:\nolinkurl{10.1073/pnas.1912789117}}


\bibitem[Efron(2011)]%
        {efron-2011-tweedie}
\bibfield{author}{\bibinfo{person}{Bradley Efron}.}
  \bibinfo{year}{2011}\natexlab{}.
\newblock \showarticletitle{Tweedie's Formula and Selection Bias}.
\newblock \bibinfo{journal}{\emph{J. Amer. Statist. Assoc.}}
  \bibinfo{volume}{106}, \bibinfo{number}{496} (\bibinfo{year}{2011}),
  \bibinfo{pages}{1602--1614}.
\newblock
\href{https://doi.org/10.1198/jasa.2011.tm11181}{doi:\nolinkurl{10.1198/jasa.2011.tm11181}}


\bibitem[Epstein et~al\mbox{.}(2025)]%
        {epstein-2025-attentionfactors}
\bibfield{author}{\bibinfo{person}{Elliot~L. Epstein}, \bibinfo{person}{Rose
  Wang}, \bibinfo{person}{Jaewon Choi}, {and} \bibinfo{person}{Markus Pelger}.}
  \bibinfo{year}{2025}\natexlab{}.
\newblock \showarticletitle{Attention Factors for Statistical Arbitrage}. In
  \bibinfo{booktitle}{\emph{Proceedings of the 6th ACM International Conference
  on AI in Finance}} \emph{(\bibinfo{series}{ICAIF '25})}.
  \bibinfo{publisher}{ACM}, \bibinfo{address}{New York, NY, USA},
  \bibinfo{pages}{396--404}.
\newblock
\href{https://doi.org/10.1145/3768292.3770398}{doi:\nolinkurl{10.1145/3768292.3770398}}


\bibitem[Fama and French(1993)]%
        {fama-french-1993}
\bibfield{author}{\bibinfo{person}{Eugene~F. Fama} {and}
  \bibinfo{person}{Kenneth~R. French}.} \bibinfo{year}{1993}\natexlab{}.
\newblock \showarticletitle{Common Risk Factors in the Returns on Stocks and
  Bonds}.
\newblock \bibinfo{journal}{\emph{Journal of Financial Economics}}
  \bibinfo{volume}{33}, \bibinfo{number}{1} (\bibinfo{year}{1993}),
  \bibinfo{pages}{3--56}.
\newblock
\href{https://doi.org/10.1016/0304-405X(93)90023-5}{doi:\nolinkurl{10.1016/0304-405X(93)90023-5}}


\bibitem[Gatev et~al\mbox{.}(2006)]%
        {gatev-2006-pairs}
\bibfield{author}{\bibinfo{person}{Evan Gatev}, \bibinfo{person}{William~N.
  Goetzmann}, {and} \bibinfo{person}{K.~Geert Rouwenhorst}.}
  \bibinfo{year}{2006}\natexlab{}.
\newblock \showarticletitle{Pairs Trading: Performance of a Relative-Value
  Arbitrage Rule}.
\newblock \bibinfo{journal}{\emph{The Review of Financial Studies}}
  \bibinfo{volume}{19}, \bibinfo{number}{3} (\bibinfo{year}{2006}),
  \bibinfo{pages}{797--827}.
\newblock
\href{https://doi.org/10.1093/rfs/hhj020}{doi:\nolinkurl{10.1093/rfs/hhj020}}


\bibitem[Geweke(1977)]%
        {geweke-1977-dfm}
\bibfield{author}{\bibinfo{person}{John Geweke}.}
  \bibinfo{year}{1977}\natexlab{}.
\newblock \showarticletitle{The Dynamic Factor Analysis of Economic Time
  Series}.
\newblock In \bibinfo{booktitle}{\emph{Latent Variables in Socio-Economic
  Models}}, \bibfield{editor}{\bibinfo{person}{Dennis~J. Aigner} {and}
  \bibinfo{person}{Arthur~S. Goldberger}} (Eds.).
  \bibinfo{publisher}{North-Holland}, \bibinfo{address}{Amsterdam},
  \bibinfo{pages}{365--383}.
\newblock


\bibitem[Giannone et~al\mbox{.}(2008)]%
        {giannone-2008-nowcasting}
\bibfield{author}{\bibinfo{person}{Domenico Giannone},
  \bibinfo{person}{Lucrezia Reichlin}, {and} \bibinfo{person}{David Small}.}
  \bibinfo{year}{2008}\natexlab{}.
\newblock \showarticletitle{Nowcasting: The Real-Time Informational Content of
  Macroeconomic Data}.
\newblock \bibinfo{journal}{\emph{Journal of Monetary Economics}}
  \bibinfo{volume}{55}, \bibinfo{number}{4} (\bibinfo{year}{2008}),
  \bibinfo{pages}{665--676}.
\newblock


\bibitem[Gneiting et~al\mbox{.}(2007)]%
        {gneiting-2007-sharpness}
\bibfield{author}{\bibinfo{person}{Tilmann Gneiting}, \bibinfo{person}{Fadoua
  Balabdaoui}, {and} \bibinfo{person}{Adrian~E. Raftery}.}
  \bibinfo{year}{2007}\natexlab{}.
\newblock \showarticletitle{Probabilistic forecasts, calibration and
  sharpness}.
\newblock \bibinfo{journal}{\emph{Journal of the Royal Statistical Society:
  Series B}} \bibinfo{volume}{69}, \bibinfo{number}{2} (\bibinfo{year}{2007}),
  \bibinfo{pages}{243--268}.
\newblock
\href{https://doi.org/10.1111/j.1467-9868.2007.00587.x}{doi:\nolinkurl{10.1111/j.1467-9868.2007.00587.x}}


\bibitem[Gu and Dao(2023)]%
        {gu-2023-mamba}
\bibfield{author}{\bibinfo{person}{Albert Gu} {and} \bibinfo{person}{Tri Dao}.}
  \bibinfo{year}{2023}\natexlab{}.
\newblock \showarticletitle{Mamba: Linear-Time Sequence Modeling with Selective
  State Spaces}.
\newblock \bibinfo{journal}{\emph{arXiv preprint arXiv:2312.00752}}
  (\bibinfo{year}{2023}).
\newblock


\bibitem[Guijarro-Ordonez et~al\mbox{.}(2021)]%
        {guijarro-2021-deepstatarb}
\bibfield{author}{\bibinfo{person}{Jorge Guijarro-Ordonez},
  \bibinfo{person}{Markus Pelger}, {and} \bibinfo{person}{Greg Zanotti}.}
  \bibinfo{year}{2021}\natexlab{}.
\newblock \bibinfo{title}{Deep Learning Statistical Arbitrage}.
\newblock
\showeprint[arxiv]{2106.04028}~[cs.LG]


\bibitem[Hamilton(1989)]%
        {hamilton-1989-regime}
\bibfield{author}{\bibinfo{person}{James~D. Hamilton}.}
  \bibinfo{year}{1989}\natexlab{}.
\newblock \showarticletitle{A New Approach to the Economic Analysis of
  Nonstationary Time Series and the Business Cycle}.
\newblock \bibinfo{journal}{\emph{Econometrica}} \bibinfo{volume}{57},
  \bibinfo{number}{2} (\bibinfo{year}{1989}), \bibinfo{pages}{357--384}.
\newblock
\href{https://doi.org/10.2307/1912559}{doi:\nolinkurl{10.2307/1912559}}


\bibitem[Hollmann et~al\mbox{.}(2023)]%
        {hollmann-2023-tabpfn}
\bibfield{author}{\bibinfo{person}{Noah Hollmann}, \bibinfo{person}{Samuel
  M{\"u}ller}, \bibinfo{person}{Katharina Eggensperger}, {and}
  \bibinfo{person}{Frank Hutter}.} \bibinfo{year}{2023}\natexlab{}.
\newblock \showarticletitle{{TabPFN}: A Transformer That Solves Small Tabular
  Classification Problems in a Second}. In
  \bibinfo{booktitle}{\emph{International Conference on Learning
  Representations}}.
\newblock


\bibitem[Hollmann et~al\mbox{.}(2025)]%
        {hollmann-2025-tabpfn}
\bibfield{author}{\bibinfo{person}{Noah Hollmann}, \bibinfo{person}{Samuel
  M{\"u}ller}, \bibinfo{person}{Lennart Purucker}, \bibinfo{person}{Arjun
  Krishnakumar}, \bibinfo{person}{Max K{\"o}rfer}, \bibinfo{person}{Shi~Bin
  Hoo}, \bibinfo{person}{Robin~Tibor Schirrmeister}, {and}
  \bibinfo{person}{Frank Hutter}.} \bibinfo{year}{2025}\natexlab{}.
\newblock \showarticletitle{Accurate Predictions on Small Data with a Tabular
  Foundation Model}.
\newblock \bibinfo{journal}{\emph{Nature}}  \bibinfo{volume}{637}
  (\bibinfo{year}{2025}), \bibinfo{pages}{319--326}.
\newblock
\href{https://doi.org/10.1038/s41586-024-08328-6}{doi:\nolinkurl{10.1038/s41586-024-08328-6}}


\bibitem[Kalman(1960)]%
        {kalman-1960-filter}
\bibfield{author}{\bibinfo{person}{Rudolf~E. Kalman}.}
  \bibinfo{year}{1960}\natexlab{}.
\newblock \showarticletitle{A New Approach to Linear Filtering and Prediction
  Problems}.
\newblock \bibinfo{journal}{\emph{Journal of Basic Engineering}}
  \bibinfo{volume}{82}, \bibinfo{number}{1} (\bibinfo{year}{1960}),
  \bibinfo{pages}{35--45}.
\newblock


\bibitem[Kong and Das(2026)]%
        {kong-2026-tabfm}
\bibfield{author}{\bibinfo{person}{Weihao Kong} {and}
  \bibinfo{person}{Abhimanyu Das}.} \bibinfo{year}{2026}\natexlab{}.
\newblock \bibinfo{title}{Introducing {TabFM}: A Zero-Shot Foundation Model for
  Tabular Data}.
\newblock \bibinfo{howpublished}{Google Research Blog}.
\newblock
\urldef\tempurl%
\url{https://research.google/blog/introducing-tabfm-a-zero-shot-foundation-model-for-tabular-data/}
\showURL{%
\tempurl}


\bibitem[Krishnan et~al\mbox{.}(2015)]%
        {krishnan-2015-dkf}
\bibfield{author}{\bibinfo{person}{Rahul~G. Krishnan}, \bibinfo{person}{Uri
  Shalit}, {and} \bibinfo{person}{David Sontag}.}
  \bibinfo{year}{2015}\natexlab{}.
\newblock \showarticletitle{Deep Kalman Filters}.
\newblock \bibinfo{journal}{\emph{arXiv preprint arXiv:1511.05121}}
  (\bibinfo{year}{2015}).
\newblock


\bibitem[Lopez-Paz and Oquab(2017)]%
        {lopez-paz-2017-c2st}
\bibfield{author}{\bibinfo{person}{David Lopez-Paz} {and}
  \bibinfo{person}{Maxime Oquab}.} \bibinfo{year}{2017}\natexlab{}.
\newblock \showarticletitle{Revisiting Classifier Two-Sample Tests}. In
  \bibinfo{booktitle}{\emph{International Conference on Learning
  Representations}}.
\newblock


\bibitem[Marino et~al\mbox{.}(2018)]%
        {marino-2018-avf}
\bibfield{author}{\bibinfo{person}{Joseph Marino}, \bibinfo{person}{Milan
  Cvitkovic}, {and} \bibinfo{person}{Yisong Yue}.}
  \bibinfo{year}{2018}\natexlab{}.
\newblock \showarticletitle{A General Method for Amortizing Variational
  Filtering}. In \bibinfo{booktitle}{\emph{Advances in Neural Information
  Processing Systems 31}}.
\newblock


\bibitem[M{\"u}ller et~al\mbox{.}(2022)]%
        {muller-2022-pfn}
\bibfield{author}{\bibinfo{person}{Samuel M{\"u}ller}, \bibinfo{person}{Noah
  Hollmann}, \bibinfo{person}{Sebastian~Pineda Arango}, \bibinfo{person}{Josif
  Grabocka}, {and} \bibinfo{person}{Frank Hutter}.}
  \bibinfo{year}{2022}\natexlab{}.
\newblock \showarticletitle{Transformers Can Do {Bayesian} Inference}. In
  \bibinfo{booktitle}{\emph{International Conference on Learning
  Representations}}.
\newblock


\bibitem[{National Bureau of Economic Research}(2026)]%
        {nber-bcdc}
\bibfield{author}{\bibinfo{person}{{National Bureau of Economic Research}}.}
  \bibinfo{year}{2026}\natexlab{}.
\newblock \bibinfo{title}{Business Cycle Dating}.
\newblock
  \bibinfo{howpublished}{\url{https://www.nber.org/research/business-cycle-dating}}.
\newblock
\shownote{Accessed August 2026}.
\newblock


\bibitem[Papamakarios and Murray(2016)]%
        {papamakarios-2016-npe}
\bibfield{author}{\bibinfo{person}{George Papamakarios} {and}
  \bibinfo{person}{Iain Murray}.} \bibinfo{year}{2016}\natexlab{}.
\newblock \showarticletitle{Fast $\epsilon$-free Inference of Simulation Models
  with {Bayesian} Conditional Density Estimation}. In
  \bibinfo{booktitle}{\emph{Advances in Neural Information Processing Systems
  29}}. \bibinfo{pages}{1028--1036}.
\newblock


\bibitem[Qu et~al\mbox{.}(2025)]%
        {qu-2025-tabicl}
\bibfield{author}{\bibinfo{person}{Jingang Qu}, \bibinfo{person}{David
  Holzm{\"u}ller}, \bibinfo{person}{Ga{\"e}l Varoquaux}, {and}
  \bibinfo{person}{Marine {Le Morvan}}.} \bibinfo{year}{2025}\natexlab{}.
\newblock \showarticletitle{{TabICL}: A Tabular Foundation Model for In-Context
  Learning on Large Data}. In \bibinfo{booktitle}{\emph{Proceedings of the 42nd
  International Conference on Machine Learning}}.
\newblock


\bibitem[Romano et~al\mbox{.}(2019)]%
        {romano-2019-cqr}
\bibfield{author}{\bibinfo{person}{Yaniv Romano}, \bibinfo{person}{Evan
  Patterson}, {and} \bibinfo{person}{Emmanuel~J. Cand\`es}.}
  \bibinfo{year}{2019}\natexlab{}.
\newblock \showarticletitle{Conformalized Quantile Regression}. In
  \bibinfo{booktitle}{\emph{Advances in Neural Information Processing Systems
  32}}.
\newblock


\bibitem[Smith and Winkler(2006)]%
        {smith-2006-optimizer}
\bibfield{author}{\bibinfo{person}{James~E. Smith} {and}
  \bibinfo{person}{Robert~L. Winkler}.} \bibinfo{year}{2006}\natexlab{}.
\newblock \showarticletitle{The Optimizer's Curse: Skepticism and Postdecision
  Surprise in Decision Analysis}.
\newblock \bibinfo{journal}{\emph{Management Science}} \bibinfo{volume}{52},
  \bibinfo{number}{3} (\bibinfo{year}{2006}), \bibinfo{pages}{311--322}.
\newblock
\href{https://doi.org/10.1287/mnsc.1050.0451}{doi:\nolinkurl{10.1287/mnsc.1050.0451}}


\bibitem[Stock and Watson(2002)]%
        {stock-watson-2002-dfm}
\bibfield{author}{\bibinfo{person}{James~H. Stock} {and}
  \bibinfo{person}{Mark~W. Watson}.} \bibinfo{year}{2002}\natexlab{}.
\newblock \showarticletitle{Forecasting Using Principal Components from a Large
  Number of Predictors}.
\newblock \bibinfo{journal}{\emph{J. Amer. Statist. Assoc.}}
  \bibinfo{volume}{97}, \bibinfo{number}{460} (\bibinfo{year}{2002}),
  \bibinfo{pages}{1167--1179}.
\newblock
\href{https://doi.org/10.1198/016214502388618960}{doi:\nolinkurl{10.1198/016214502388618960}}


\bibitem[Talts et~al\mbox{.}(2018)]%
        {talts-2018-sbc}
\bibfield{author}{\bibinfo{person}{Sean Talts}, \bibinfo{person}{Michael
  Betancourt}, \bibinfo{person}{Daniel Simpson}, \bibinfo{person}{Aki Vehtari},
  {and} \bibinfo{person}{Andrew Gelman}.} \bibinfo{year}{2018}\natexlab{}.
\newblock \bibinfo{title}{Validating {Bayesian} Inference Algorithms with
  Simulation-Based Calibration}.
\newblock
\showeprint[arxiv]{1804.06788}~[stat.ME]


\end{thebibliography}

\appendix
\section{The Generative Model}
\label{app:dgp}

The scale, level, and innovation law of \eqref{eq:sou} evolve as
\begin{equation}
\begin{aligned}
\log \sigma_{i,t} &= \phi \log \sigma_{i,t-1} + \eta\,\nu_{i,t},
  &
m_{i,t} &= m_{i,t-1} + \delta\,\xi_{i,t},\\[2pt]
\zeta_{i,t} &= \bigl(e^{\gamma\,u_{i,t}} - \mathbb{E}[e^{\gamma u}]
                      \bigr)\big/ \operatorname{sd}[e^{\gamma u}],
\end{aligned}
  \label{eq:sv}
\end{equation}
with $\nu, u, \xi \sim \mathcal{N}(0,1)$. Table~\ref{tab:components}
lists the parameters of each component and how each is fixed. Only
the innovation law and the scale are fitted to data, by moment
matching on the training split's real windows and never on the
two-sample test; the kernel carries the two speeds, $\eta$ is fixed
so that $\log\sigma$ has stationary standard deviation $0.5$, and
the level step $\delta$ is held at a fixed default. The two-sample test constrains the innovation law and the scale but
not the kernel, so the number of regimes is a modeling choice, two being the smallest set that separates a
reversion the trade can use from one it cannot. The level is untested, every candidate being simulated with $m
\equiv 0$; a level $g(X)$ learned from characteristics by gradient boosting on
slow window statistics and the six fundamentals, fitted causally,
does not outperform a $250$-day trailing mean of the spread on
held-out data, and is retained only as the pre-registered anchor of
the overshoot test in Section~\ref{sec:experiment:trading}. The
injected positive controls of Section~\ref{sec:experiment} add a
trend away from the level.

\begin{table}[t]  \caption{Two-sample AUC between real and simulated residual windows,
    mean $\pm$ SE over $5$ seeds.}
  \label{tab:admission}
  \footnotesize
  \setlength{\tabcolsep}{2.0pt}
  \begin{tabular}{l cc l}
    \toprule
    innovation law & crypto & KRX & verdict \\
    \midrule
    Gaussian, constant $\sigma$ & $0.903 \pm .002$ & $0.904 \pm .001$ & fail \\
    Student-$t$ & $0.809 \pm .002$ & $0.788 \pm .002$ & fail \\
    lognormal & $0.890 \pm .001$ & $0.885 \pm .001$ & fail \\
    stochastic vol (SV) & $0.744 \pm .004$ & $0.712 \pm .002$ & fail/pass \\
    \textbf{SV + lognormal} & $\mathbf{0.686 \pm .003}$ & $\mathbf{0.678 \pm .002}$ & \textbf{pass} \\
    random walk, Gaussian & $0.895 \pm .001$ & $0.898 \pm .001$ & fail \\
    \bottomrule
  \end{tabular}
\end{table}

\begin{table}[t]
  \caption{Components of the generative model.}
  \label{tab:components}
  \scriptsize
  \setlength{\tabcolsep}{3pt}
  \begin{tabular}{@{}l l l l c@{}}
    \toprule
    component & form & parameters & fixed by & tested \\
    \midrule
    kernel & OU, regime-indexed speed & $\kappa(\textup{C}), \kappa(\textup{D})$ & prior & no \\
    innovations & centered lognormal & $\gamma$ & moment matching & yes \\    scale & AR(1) log-volatility & $\phi$, $\eta$ & $\phi$ matched; $\eta$ fixed & yes \\    level & random walk, step $\delta$ & $\delta$ & fixed & no \\
    \bottomrule
  \end{tabular}
\end{table}

\emph{The two-sample test.} For each candidate innovation law a
classifier is trained to distinguish simulated residual windows from
real ones and its AUC recorded (Table~\ref{tab:admission}); a
candidate passes when its AUC plus two standard errors lies below
$0.75$.
\begin{description}\setlength{\itemsep}{0pt}\setlength{\parsep}{0pt}
\item[Windows] real: one fixed set of cross-sectionally demeaned
  cumulative residuals over the full sample (the PCA and AFM series
  are not re-tested); simulated: five seeds, which give the standard
  error.
\item[Classifier] gradient boosting on scale-invariant window
  statistics: tail weight, skewness, autocorrelations of the increments
  and of their absolute values, zero crossings, range, half-window
  volatility ratio.
\item[Controls] same-process windows score $0.50$ and a wrong process
  $0.90$; the kernel moves the AUC by at most $0.014$, so the
  statistics constrain the innovation law, not the kernel.
\item[Candidates] fifteen: the laws of Table~\ref{tab:admission} and
  their combinations, plus random walks under the same laws as
  controls without mean reversion; the table reports one
  representative per feature.
\end{description}
Student-$t$ innovations score $0.809$/$0.788$ and worsen the fit
when combined with stochastic volatility, so the data are better matched by a day-to-day varying scale with
asymmetric innovations than by a fixed fat-tailed law. The bar of
$0.75$, halfway between chance and perfect separation, was
registered before any run as a convention, not derived; at $0.678$
the selected process remains distinguishable from the data, and the
posterior claims on the KRX portfolio should be interpreted in light
of that residual discrepancy.

\section{Experiment Details}
\label{app:details}

\begin{table}[t]
  \caption{Protocol and configuration, fixed in advance.}
  \label{tab:setup}
  \scriptsize  \setlength{\tabcolsep}{2pt}
  \setlength{\aboverulesep}{1pt}\setlength{\belowrulesep}{1pt}
  \renewcommand{\arraystretch}{0.92}
  \begin{tabular}{@{}>{\raggedright\arraybackslash}p{0.25\columnwidth}
                  >{\raggedright\arraybackslash}p{0.73\columnwidth}@{}}
    \toprule
    universe & top-$200$ KRX by median traded value; $120$ names with
      complete history, $2010$--$2026$ ($4{,}072$ d) \\
    characteristics & ROE, D/E, net margin, revenue growth, B/P, E/P
      (DART, point-in-time, from $2016$) \\
    split & $2018$-$04$; the first $30\%$ of each period trains \\    window, hold & $L = 60$, $H = 20$ (non-overlapping); $\tau^\ast = 60$\,d \\
    residuals & PCA, $K = 3$ on $252$ d; FF3 on $[1, \text{size},
      \text{B/P}]$ \\
    sizing & $64$ posterior draws; $d \in [-3, 3]$; dollar-neutral at
      unit gross \\
    costs & $5$\,bp per unit of turnover; $1$\,bp per day on the short
      notional \\
    benchmark, nulls & random signal held $H$ d; vol-matched random
      walk; $10$d/$120$d switching \\
    \midrule    encoder, training & GRU-$64$; five-component MDN; Adam, $4{,}000$ steps, batch $512$, rate
      $3{\times}10^{-3}$ cosine; $60{,}000$ draws \\
    half-life prior & $\log U(3, 400)$\,d; per-stock centres at test \\    $(\phi, \gamma)$, training split & $(0.09, 0.08)$ in $2010$--$18$;
      $(0.28, 0.12)$ in $2018$--$26$ \\
    two-state chain & half-lives $10$\,d (C) and $120$\,d (D) \\
    \midrule
    AFM & $K = 5$; annual refits; $40$ epochs \\
    panel form & $2$ blocks, width $64$, $4$ heads; $30$k panels, $3$k
      steps, batch $16$, rate $3{\times}10^{-4}$ \\
    \bottomrule
  \end{tabular}
\end{table}

\emph{Data.} Both universe filters use end-of-sample information
(liquidity ranking at the sample end; delisted stocks dropped). The
six characteristics are point-in-time: each becomes available on the
first trading day after disclosure, flows are trailing-twelve-month,
capitalization is split-consistent, and a missing stock-day receives
a neutral rank.

\emph{Definitions.}
\begin{description}\setlength{\itemsep}{0pt}\setlength{\parsep}{0pt}
\item[Weights and costs] defined on the residual portfolios of
  \eqref{eq:rowstep} \cite{epstein-2025-attentionfactors}; the implied
  factor-hedge legs are not separately traded or costed.
\item[Turnover] the mean daily absolute change in these weights as a
  fraction of gross.
\item[$\Delta$SR] the annualized Sharpe ratio of the daily return
  difference between a strategy and the random benchmark (bootstrap SE
  $0.41$--$0.42$); the benchmark redraws its random positions every $H$ days at the
  same gross, hence its higher turnover.
\item[Not modeled] the Korean transaction tax and short-sale
  restrictions in force during part of the window; levels are
  therefore not implementable returns.
\end{description}

\emph{Estimation notes.}
\begin{description}\setlength{\itemsep}{0pt}\setlength{\parsep}{0pt}
\item[FF3] needs size and B/P on the same stock-day, with joint
  coverage only from $2017$, so part of its weakness, in the earlier
  period especially, reflects limited coverage.
\item[Conformal layer] widths are multiplicative per quantile;
  the outcome-fitted methods fit on the first $80\%$ of the training split
  and calibrate on the remaining $20\%$ after an $L$-day gap,
  simulator methods on the full training split.
\item[Shrinkage control] keeps the point estimate's positions and
  rescales only its forecast by one constant fixed on the training
  split; its KRX run uses a panel rebuilt under the same rule from a
  second vendor's adjusted prices, on which Table~\ref{tab:honesty}'s
  ordering also holds.
\end{description}

\end{document}